\documentclass{aa}
\usepackage{natbib,amssymb,graphicx,graphics}
\newcommand{\excs}{\extracolsep{\fill}}
\newcommand{\bpw}{\object{$\beta\:$Pictoris}}
\newcommand{\bp}{\object{$\beta\:$\,Pic}}
\newcommand{\bpb}{\object{$\beta\:$\,Pic b}}

\newcommand{\msun}{M_\odot}

\begin{document}

\title{Orbital characterization of the $\beta$ Pictoris b giant planet\thanks{Based on observations collected at the European Southern Observatory, Chile (ESO programmes 072.C-0624, 384.C-0207, 084.C-0739, 284.C-5057, 385.C-0132, 086.C-0341)}}
\titlerunning{$\beta$ Pictoris b orbital characterization}

\author{G. Chauvin\inst{1,2}
	\and A.-M. Lagrange\inst{1}
	\and H. Beust\inst{1}
	\and M. Bonnefoy \inst{2}
	\and A. Boccaletti \inst{3}
 \and 
  D. Apai \inst{4}
 \and 
  F.~Allard \inst{5}
  \and
  D.~Ehrenreich \inst{1}
  \and
  J. H. V. ~Girard \inst{6}
  \and
  D.~Mouillet \inst{1}
  \and
  D.~Rouan \inst{3} 
}
\institute{
$^{1}$UJF-Grenoble 1 / CNRS-INSU, Institut de Plan\'etologie et d'Astrophysique de Grenoble (IPAG) UMR 5274, Grenoble, F-38041, France\\
$^{2}$Max Planck Institute for Astronomy, K\"onigstuhl 17, D-69117 Heidelberg, Germany\\
$^{3}$LESIA, Observatoire de Paris Meudon, 5 pl. J. Janssen, 92195 Meudon, France\\ 
$^{4}$ Department of Astronomy and Department of Planetary Sciences, The University of Arizona, 933 N Cherry Avenue, Tucson, AZ 85718
$^{5}$ Centre de Recherche Astronomique de Lyon, 46 all\'ee d'Italie, 69364 Lyon cedex 7, France\\
$^{7}$  European Southern Observatory, Casilla 19001, Santiago 19, Chile 	\\
}

\offprints{G. Chauvin}
\date{Received ; accepted }

  \abstract 
  {In June 2010, we confirmed the existence of a giant planet in the
    disk of the young star \bpw\, located between 8~AU and 15~AU from the
    star. This young planet offers the rare opportunity to monitor a
    large fraction of the orbit using the imaging technique over a
    reasonably short timescale. It also offers the opportunity to study its
    atmospheric properties using spectroscopy and multi-band photometry, and possibly
    derive its dynamical mass by combining imaging with radial velocity data to
    set tight constraints on giant planet formation theories.}
    {We aim to measure the evolution of the planet's position relative
      to the star \bpw\ to determine the planetary orbital
      properties. Our ultimate goal is to relate both the planetary
      orbital configuration and physical properties to either the disk
      structure or the cometary activity observed for decades in the
      \bpw\ system.}
   {Using the NAOS-CONICA adaptive-optics instrument (NACO) at the
     Very Large Telescope (VLT), we obtained repeated follow-up images
     of the \bpw\ system in the $K_s$ and $L~\!'$ filters at
       four new epochs in 2010 and 2011. Complementing these data with previous
     measurements, we conduct a homogeneous analysis,
     which covers more than eight~yrs, to accurately monitor the \bpw~b
     position relative to the star. }
   {On the basis of the evolution of the planet's relative position
     with time, we derive the best-fit orbital solutions for our
     measurements. More reliable results are found with  a Markov-chain Monte Carlo
     approach. The solutions favor a
     low-eccentricity orbit $e\la0.17$, with semi-major axis in the
     range 8--9\,AU corresponding to orbital periods of
     17--21\,yrs. Our solutions favor a highly inclined solution with
     a peak around $i=88.5\pm1.7^o$, and a longitude of ascending node
     tightly constrained at $\Omega=-147.5\pm1.5\degr$. These results
     indicate that the orbital plane of the planet is likely to be
     above the midplane of the main disk, and compatible with the warp
     component of the disk being tilted between 3.5~deg and
     4.0~deg. This suggests that the planet plays a key role in the
     origin of the inner warped-disk morphology of the
     \bp\ disk. Finally, these orbital parameters are consistent with
     the hypothesis that the planet is responsible for the
     transit-like event observed in November 1981, and also linked to
     the cometary activity observed in the \bp\ system.}

    {}
   \keywords{Techniques: adaptive optics, high angular resolution; Astrometry; Methods: observational, data analysis; Stars: planetary systems}
   \maketitle

\section{Introduction}

\begin{figure*}[t]
\includegraphics[width=14cm]{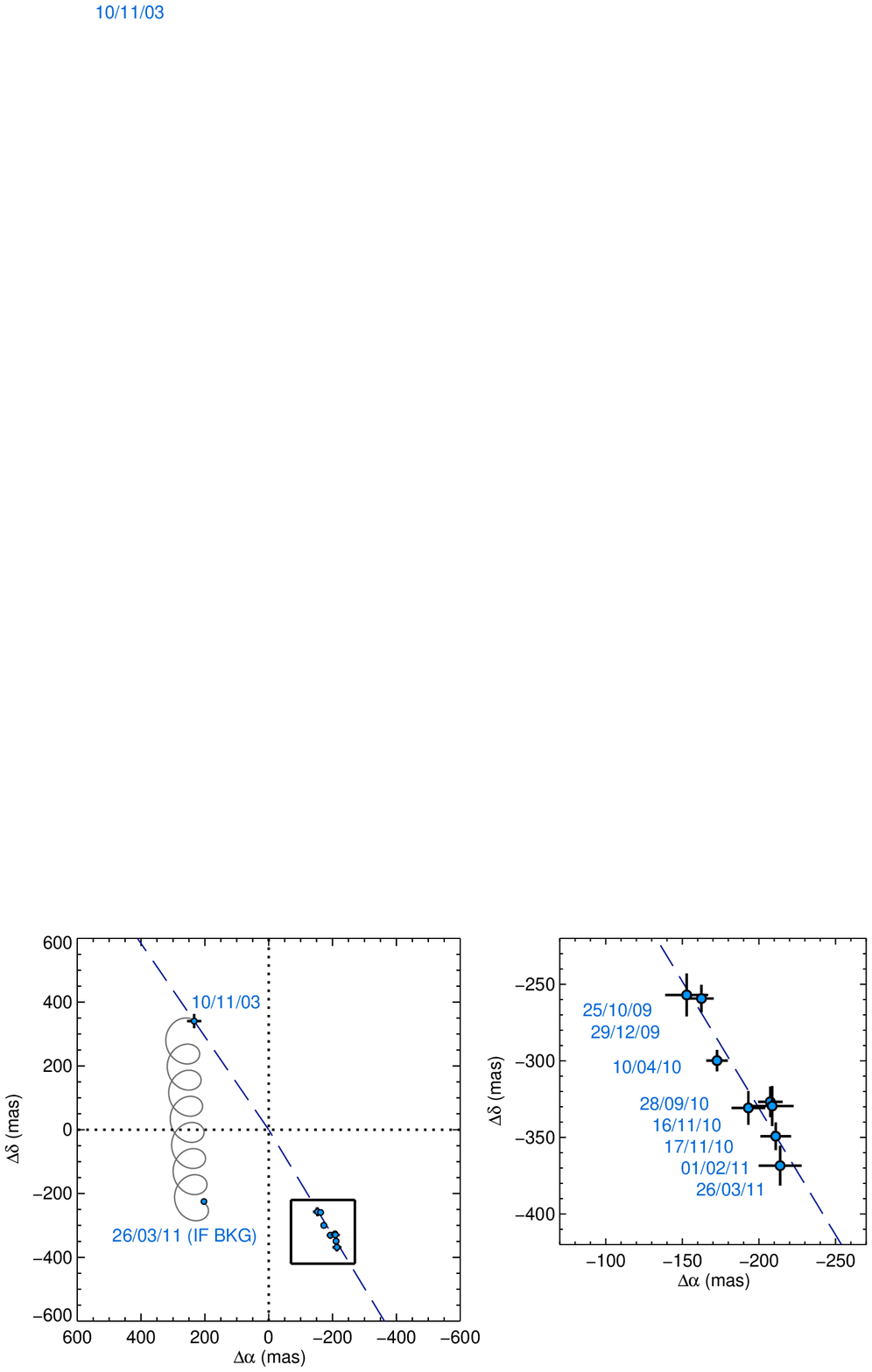}
\begin{centering}
\caption{Left: Astrometric positions of \bpb\ relative from A used in
  the present work to characterize the orbital parameters of the
  \bpb\ planet. The predictions for the \bpb\ in case of a background
  source are reported in gray from November, 10, 2003 to March, 26
  2011. Two linear fits, passing through the origin $(\Delta\alpha,
  \Delta\delta)=(0,0)$, for the 2003 NE data points (at 34.25~deg) and
  the 2009, 2010 and 2011 SW measurements (at 211.15~deg) are
  indicated by the two blue dashed lines. A zoomed view of the most recent
  astrometric observations over 2010 and 2011 is presented on the
  right.}
\label{fig:astro}
\end{centering}
\end{figure*}

In the context of exoplanetary science, the direct imaging technique
offers a unique observing window to explore the frequency and the
properties of the extrasolar giant planet (EGP) population in large
orbits ($\geq 5$~AU). The recent discoveries of massive planetary mass
companions around stars and brown dwarfs
\citep{cha04,cha05,laf08,kal08,mar08,lag09,mar10} have shown that
core-accretion alone cannot explain the formation of all
imaged giant planets, because the core-accretion timescales become
too long and the disk surface density too low. Disk or core
fragmentation are alternative mechanisms that could even operate
in wide orbits, leading to different physical and orbital
distributions \citep{bol09,dod09,vor10}. Additional mechanisms such as
inward/outward migrations or planet-planet interactions might also
modify the EGP orbital properties \citep{cri09}. Consequently, there
is a real need to characterize the orbital properties of recently
imaged EGPs, to determine the planetary system architectures and
dynamical stabilities, and to obtain additional constraints on their
formation and evolution mechanisms. Follow-up astrometric studies have
been conducted on the HR\,8799 multiple exoplanetary system
\citep{laf09,mar10,ber11,sou11}. Inclined orbital solutions for the b
and d planets, and a 1:2:4 resonance for the planets c, d, and e, confirm
stable configurations supported by dynamical simulations
\citep{rei09,fab10}. Owing to its even smaller semi-major axis, the
  planet \bpw~b (hereafter \bpb) offers a rare opportunity to
  constrain both the planetary orbital and physical properties (Lagrange et al. 2010,
  Currie et al. 2011). This massive giant planet orbits at 8--15~AU
  around the young star $\beta$ Pictoris (A5V, $19.44\pm0.05$~pc,
  $12_{-4}^{+8}$~Myr), which has been studied for almost three decades because of its
  emblematic debris disk (see Vidal-Madjar et al. 1998 for
  a review). So-called hot start theoretical models predict a mass of
  $9\pm3$~M$_{\rm{Jup}}$ and $T_{\rm{eff}}=1700\pm300$~K (Quanz et
  al. 2010, Bonnefoy et al. 2011). Characterizing the planetary
  properties offers a rare opportunity to directly investigate the
  planet-disk interactions, particularly the role played by the planet
  in the formation of the the inner warped component of the
  \bp\ circumstellar disk \citep{mou97,aug01,daw11}.

Since the recovery of \bpb\ (i.e its re-discovery after its passage
behind the star), we have performed an astrometric monitoring
campaign, using NACO at the VLT. In Sect.~2, we describe
observations aquired in the years 2010 and 2011. In Sect.~3, we
present the data analysis of these new data together with previous
measurements including available astrometric calibrations (Lagrange et
al. 2009, 2010; Bonnefoy et al. 2011). Our main objective is to
homogeneously analyze the astrometric position of \bpb\ relative to the
central star to accurately constrain its orbital properties. In
Sect.~4, we report the most probable solutions for the planet's orbit
that best fits our astrometric measurements, using two orbital fitting
methods, a least-square Levenberg-Marquardt algorithm and a
Markov-chain Monte Carlo approach. Finally, in Sect~5, we discuss the
consequences of our results in the context of previous astrometric
studies, and their implications for a possible connection with
the past transiting event observed in 1981, the disk structures, and
the cometary activity of the \bp\ system.

\section{Observations}

To monitor the \bpb\ astrometry, we used the NACO high contrast
adaptive optics (AO) imager of the VLT, equipped withthe NAOS AO
system \citep{rou02}, and the near-infrared imaging camera CONICA
\citep{len02}. The follow-up observations were obtained at five
different epochs between September 2010 and March 2011, using the
angular differential imaging (ADI) mode of NACO. For accurate
astrometry, two observing set-ups were used, the $L~\!'$ filter with
the L27 camera and the $K_s$ filter with the S27 camera. The NACO
detector cube mode was in addition used for frame selection.  A
classical dithering sequence was used with the repetition of five
offset positions to properly remove the sky contribution. In the end,
the typical observing sequence represented a total of 200--250 cubes,
i.e a total integration time of 35--50~min for an observing sequence
of 1--1.5~hrs on target. The parallactic angles at the start and the
end of our observations are reported in Table~\ref{tab:obs}, together
with the exposure time (DIT), the number of frame per cube (NDIT), and
the number of cubes (Nexp) for each epoch. Typical exposure times of
0.15~s and 0.2~s were used in the $K_s$ and $L~\!'$-filters,
respectively, to saturate the point spread function (PSF) core by a
factor of 100 (a few pixels in radius) to improve the dynamics of our
images. The observing sequences were executed to optimize both the field
rotation and the position of the secondary mirror diffraction spikes
relative to the companion, except in March 26, 2011 when the data were
obtained in service observing mode. Two sequences of non-saturated
PSFs were acquired using a neutral density filter at the beginning and the
end of each observing sequence to monitor the image quality. These data also
served for the calibration of the relative photometric and astrometric
measurements of $\beta$ Pic~b.  During the different observing sequences
obtained in visitor and service queue modes at ESO, the atmospheric
conditions were stable, with seeing and coherence times of 0.6--1.0\,\arcsec\ and 3--6~ms , respectively. The same
  astrometric field was observed within a week of each follow-up
  observation of $\beta$ Pic (see below).

\begin{table}[t]
\caption{Obs log of new 2010--2011 VLT/NACO observations}             
\label{tab:obs}
\centering
\begin{tabular*}{\columnwidth}{@{\excs}llllll}     
\hline\hline\noalign{\smallskip}
UT Date    &    Filter/Camera   &    DIT   &   NDIT   & $\rm{N}_{\rm{exp}}$   & $\theta_s;\theta_e$ \\
           &                    &    (s)   &      &    & (deg);(deg) \\
\noalign{\smallskip}\hline\noalign{\smallskip}
28/09/10   &    $L~\!'$/L27           &    0.2   &   150    & 199                  & $-60;-9$  \\
           &    $L~\!'$-ND$_l$/L27    &    0.2   &   100    & 20                   & $-62;-61$ \\
16/11/10   &    $K_s$/S27           &    0.15  &   100    & 199                  & $-21;+14$ \\
           &    $K_s$-ND$_s$/S27    &    0.11  &   100    & 10                   & $-23;-21$ \\
17/11/10   &    $L~\!'$/L27           &    0.2   &   100    & 268                  & $-35;+32$ \\
           &    $L~\!'$-ND$_l$/L27    &    0.2   &   100    & 10                   & $-37;-35$ \\
01/02/11   &    $K_s$/S27           &    0.15  &   100    & 223                  & $-6;+36$  \\
           &    $K_s$-ND$_s$/S27    &    0.15  &   100    & 10                   & $-9;-6$ \\
26/03/11   &    $K_s$/S27           &    0.15  &   100    & 249                  & $+38;+62$  \\
           &    $K_s$-ND$_s$/S27    &    0.15  &   100    & 10                   & $-9;-6$ \\
\hline\noalign{\smallskip}
\end{tabular*}
\end{table}

\begin{table*}[t]
\caption{NACO astrometric measurements of \bpb\ relative to \bp}             
\label{tab:astro}
\centering
\renewcommand{\footnoterule}{}  
\begin{tabular*}{\textwidth}{@{\excs}llllllllllll}     
\hline\hline\noalign{\smallskip}
UT Date     & Mode          &  Platescale     & True North$^a$ & Rotator Offset$^b$          & $\Delta\alpha$&$\Delta \delta$ & Separation      &     PA         \\ 
            & Obs/Filter/Obj& (mas)           & (deg)  & (deg)               & (mas)         & (mas)          & (mas)           & (deg)          \\  
\noalign{\smallskip}\hline\noalign{\smallskip}
   10/11/03  & Field/$L~\!'$/L27 & $27.11\pm0.04$  &    $0.29\pm0.07$ &   0. & $233\pm22$    &   $341\pm22$   &   $413\pm22$    &  $34.42\pm3.52$   \\
  25/10/09   & Field/$L~\!'$/L27 & $27.11\pm0.05$  &    $-0.08\pm0.10$ &  0. & $-153\pm14$   &   $-257\pm14$  &   $299\pm14$    &  $210.74\pm2.89$ \\
  29/12/09   & ADI/$L~\!'$/L27   & $27.10\pm0.04$  &    $-0.06\pm0.08$ &  $90.46\pm0.10$ & $-163\pm9$    &   $-260\pm8$   &   $306\pm9$     &  $212.07\pm1.71$ \\
  10/04/10   & ADI/$K_s$/S27   & $27.01\pm0.04$  &    $-0.26\pm0.09$ &  $90.46\pm0.10$ & $-173\pm7$    &   $-300\pm7$   &   $346\pm7$     &  $209.93\pm1.15$ \\
  28/09/10   & ADI/$L~\!'$/L27   & $27.11\pm0.04$  &    $-0.36\pm0.11$ &  $90.47\pm0.10$ & $-193\pm11$   &   $-331\pm11$  &   $383\pm11$    &  $210.28\pm1.73$ \\
  16/11/10   & ADI/$K_s$/S27   & $27.01\pm0.05$  &    $-0.25\pm0.07$ &  $90.47\pm0.10$ & $-207\pm8$    &   $-326\pm10$  &   $387\pm8$     &  $212.41\pm1.35$ \\ 
  17/11/10   & ADI/$L~\!'$/S27   & $27.10\pm0.04$  &    $-0.25\pm0.07$ &  $90.46\pm0.10$ & $-209\pm13$   &   $-330\pm14$  &   $390\pm13$    &  $212.34\pm2.13$ \\ 
  01/02/11   & ADI/$K_s$/S27   & $27.01\pm0.04$  &    $-0.32\pm0.10$ &   $90.46\pm0.10$ & $-211\pm9$   &   $-350\pm10$  &   $408\pm9$     &  $211.13\pm1.48$ \\
  26/03/11   & ADI/$K_s$/S27   & $27.01\pm0.04$  &    $-0.35\pm0.10$ &  $90.46\pm0.10$ & $-214\pm12$   &   $-367\pm14$  &   $426\pm13$    &  $210.13\pm1.81$ \\
\hline\noalign{\smallskip}
\end{tabular*}
\begin{list}{}{}
\item[$^{\mathrm{a}}$] The orientation of true north is relative to
  the vertical of the detector, and is positive when lying to the east
  of the vertical.
\item[$^{\mathrm{b}}$] The NACO rotator offset position was
  calibrated and linked to the ESO keyword ADA.PUPILPOS according
  to the formulae $\rm{ROTOFF} = 179.44 -
  \rm{ADA.PUPILPOS}$, using various astrometric binaries observed at
  various epochs between November and December 2010. A conservative
  error of 0.10~deg was considered for this calibration.
\end{list}
\end{table*}

\section{Data analysis}

For the present study, we processed the data of the new observations
of \bpb\, obtained in September 28 2010, November 16 2010, November 17
2010, February 1st 2011, and March 26 2011. Previously archived data
including available astrometric calibrations, obtained between
November 2003 and April 2010, were also re-processed (see
Table~\ref{tab:astro} and Fig.~1 ). The most robust astrometric measurements
(in terms of observing conditions and stability) were selected at each
epoch. Data obtained on November 16, 2010 and November 17, 2010 were
both reduced to check the consistency of astrometric results obtained
in the setups $L~\!'$/L27 and $K_s$/S27 at a given epoch, as both setups were
used for the astrometric analysis. The VLT/NACO $M~\!'$ data of Currie
et al. (2011) could not be considered owing to a lack of information
about the NACO rotator offset position during the observation that could not be
recovered from the ESO keywords, and that may affect the final
absolute angular position. All data were homogeneously flat-fielded,
cleaned from bad and hot pixels, and sky-subtracted. Sub-frames of
200$\times$200 pixels were extracted to reduce the computing time of
the data processing. The frames were recentered based on a
  registration of the central star position measured by a Moffat
  fitting of the non-saturated part of the stellar PSF wing, that was
  a threshold equal to at 1\%  of the detector linearity (i.e about 60\% of the
  saturation limit).  We also automatically decided to reject
open-loop and poor-AO correction images, using a PSF encircled energy
criterion.

For PSF-subtraction of the field-tracking data of November 11, 2003
and October 25, 2009, a reference star matching the \bp\ observations
in terms of parallactic angles was used to minimize the residuals in
the final images (see Lagrange et al. 2009). For ADI data, three
different ADI-algorithms were applied to the PSF-subtraction, to
verify the consistency of the solutions obtained using the classical
ADI, smart ADI, and LOCI algorithms \citep[see][for more details on
  the algorithms]{lag10,laf07}. For smart ADI, we considered a
separation criteria of 1.0 and $1.5\times FWHM$, at the companion
separation (ranging from 11 to 16 pixels), and ten images
to compute each individual PSF. For LOCI, we considered optimization
regions of $N_A = 300\times FWHM$, the radial-to-azimuthal width ratio
$g=1$, the radial width $\Delta r=2\times FWHM$, and a separation
criteria of 1.0 and $1.5\times FWHM$. Only smart ADI results were
finally considered owing the the ability of that method to reduce the
companion self-subtraction, optimize the temporal PSF selection,
and minimize the algorithm complexity when testing the photometric
and astrometric systematics induced by the algorithm itself. We always
found consistent results within the measurement uncertainty, which are
detailed below, when using the classical ADI and LOCI algorithms.

The main difficulty in deriving the planet's position relative to the
primary star was to accurately estimate the position of a central star
with a saturated core, and the position of the low signal-to-noise
planet affected by the stellar residuals. To determine the central star
position, as for the frames recentering, a Moffat fitting of
  the non-saturated part of the stellar PSF wing (with a similar
  threshold) was used. For the planet position and flux, we used a
grid of 5000 negative fake planets scanning a three-dimensional parameter
space in (X,Y) positions (sampling of 0.02 pixels) and flux (sampling
of 10~ADU), injected one-by-one in the datacubes before
  PSF-subtraction. The datacubes were then reprocessed to derive the
  best-fit solution minimizing the residuals in the final
  subtracted-image, considering a region covering the companion ADI
  signature. The positions of the companion relative to the primary
star were finally transformed into sky coordinates using the
$\theta_1$ Ori C field observed with HST by McCaughrean \& Stauffer
(1994) (with the same set of stars TCC058, 057, 054, 034, and 026). For
ADI data, the NACO rotator offset at the start of each ADI sequence
was also calibrated and taken into account, using an astrometric
binary observed in both field-tracking and pupil-tracking mode to
estimate this offset.

The results are given in Table~\ref{tab:astro} and reported in
Fig.~1. The main contributors to the uncertainty in our astrometric
measurements are listed below. Errors 2/ and 3/ were added
quadratically, then linearly added to the systematic error in our
measurement error 1/. The errors 4/ were neglected. We know consider
the four types of error:
\begin{enumerate}
\item The first contributor is the systematic error related to the
  determination of the (saturated) star center position, which is estimated
  from the fit to the PSF wings. We used non-saturated images to
  explore this effect given the saturation factor and the Moffat
  fitting threshold. If the PSF were centro-symmetric, the center
  estimate based on either the PSF wings or the PSF core would perfectly match. Tests
  on non-saturated data show that it does not. It induces a bias of
  0.2-0.4~pixels (i.e 6-12~mas). This asymmetry is variable such that
  this offset cannot be securely calibrated and subtracted. We
  note that, for consistency purposes, we re-analyzed the 2003 data
  with exactly the same method.  Owing to the minor modification of
  the fitting pattern, this leads to slightly different values (of
  0.5~pixels) from Lagrange et al (2009).
\item A second source of error is the uncertainty in the companion
  position. Twelve negative fake planets of similar flux and
    separation at various position angles were then inserted to test
  the effect of both the stellar residuals in our measurements, and the
  measurement procedure itself. The two non-saturated PSFs were used
  to take into account their influence on the final result.  The
  typical error found was about 0.1-0.2~pixels (3-6~mas).
\item We then considered the errors related to the platescale and true
  north orientation (see Table~\ref{tab:astro}), as well as the uncertainty
  in the NACO rotator offset for ADI measurements (the error of
  0.10~deg in the rotator angular accuracy). We note that
  an accurate absolute angular calibration of the NACO detector is
  difficult to achieve at mas precision. The two main limitations are
  the significant variation of the NACO detector true north with time,
  and the uncertainty associated with the calibrators
  themselves. Comparing data from different authors using different
  calibrators may then be risky in the end. A first solution is then
  to observe the same astrometric references at each epoch to ensure
  an accurate relative astrometry (as done here). A residual angular systematic
  cannot be totally excluded but does similarly affect all
  measurements and the orbital solution (see discussion below).
\item Finally, we neglected the errors related to the distortion
  correction, detector non-linearity, differential tilt jitter, Strehl
  ratio variation, or differential refraction which were  estimated smaller than 1--2 mas; (see Fritz et al. 2010).
\end{enumerate}

\section{Orbital fit}
\begin{table}
\caption[]{Orbital solutions for \bpb: the best-fit reduced $\chi^2_r$ model
  obtained with the LSLM algorithm (top), and a typical
  ``highly probable'' orbit according to the MCMC fit (bottom).  Note
  that we do not provide error bars here, as these are assumed to be
  described by the MCMC distribution.}
\label{mle}
\begin{tabular*}{\columnwidth}{@{\excs}llllllll}
\hline\noalign{\smallskip}
$a\,$(AU) & $P\,$(yr) & $e$ & $i$ ($\degr$) & $\Omega$ ($\degr$) & 
$\omega$ ($\degr$) & $t_p\,$(yr JD) & $\chi^2_r$\\ 
\noalign{\smallskip}\hline\noalign{\smallskip}
11.2 & 28.3 & 0.16  & 88.8 & $-147.73$ & $4.0$    & 2013.3 & 0.45 \\
8.8  & 19.6 & 0.021 & 88.5 & $-148.24$ & $-115.0$ & 2006.3 & 0.56 \\
\noalign{\smallskip}\hline
\end{tabular*}
\end{table}

Our astrometric measurements reported in Table~\ref{tab:astro} show
that the position angles are almost consistent with an edge-on orbital
configuration, and are roughly consistent with the position angle of
the \bp\ circumstellar disk, which is reported to be 30.1~deg and 211.4~deg by
  Kalas \& Jewitt (1995) for the NE and SW sides respectively, but
  with no error bars for these values). To derive the best-fit solution for our
measurements, we considered the planet's inclination as a free
parameter. We assumed a Keplerian orbit described in a referential
frame $OXYZ$, where, as usual, the $XOY$ plane corresponds to the plane
of the sky and the $Z$-axis points towards the Earth. In this
formalism, the astrometric position of the planet relative to the star
is given by:

\begin{eqnarray}
x & = & \Delta\delta\;=\;r\left(\cos(\omega+v)\cos\Omega-\sin(\omega+v)
\cos i\sin\Omega\right)\qquad,\label{xmodel}\\
y & = & \Delta\alpha\;=\;r\left(\cos(\omega+v)\sin\Omega+\sin(\omega+v)
\cos i\cos\Omega\right)\qquad,\label{ymodel}
\end{eqnarray}

\noindent{where $\Omega$ is the longitude of the ascending node (measured from
North), $\omega$ is the argument of periastron, $i$ is the
inclination, $v$ is the true anomaly, and $r=a(1-e^2)/(1+e\cos v)$,
where $a$ represents for the semi-major axis and $e$ the eccentricity.}

\begin{figure}[t]
\centerline{\includegraphics[angle=-90,width=\columnwidth]{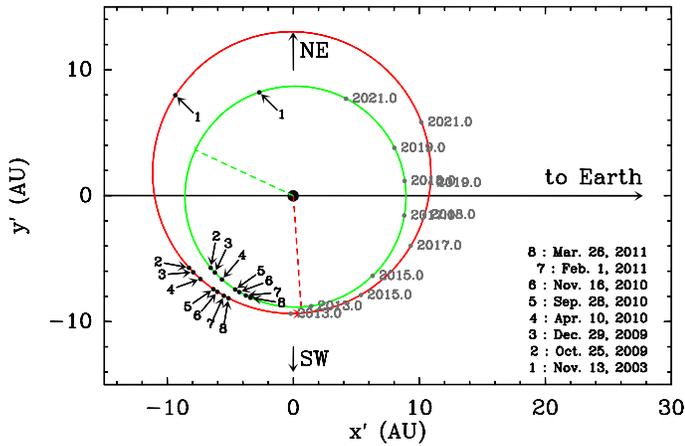}}
\caption{Plots ot the orbit of Table~\ref{mle} with their orientation
  with respect to the line of sight. The larger orbit is the best LSLM
  $\chi^2$ model and the smaller one is an example of the highly probable
  orbit obtained with the MCMC approach. In each case, the dashed line
  shows the location of the periastron.  The position of the planet at
  different observation epochs is shown as black dots along the orbit
  (projected error bars in the astrometric measurements are smaller
  than the symbol size), and predictions for the upcoming years are
  shown in gray.}
\label{orbplots}
\end{figure}

\begin{figure*}
\makebox[\textwidth]{
\includegraphics[angle=0,width=0.33\textwidth]{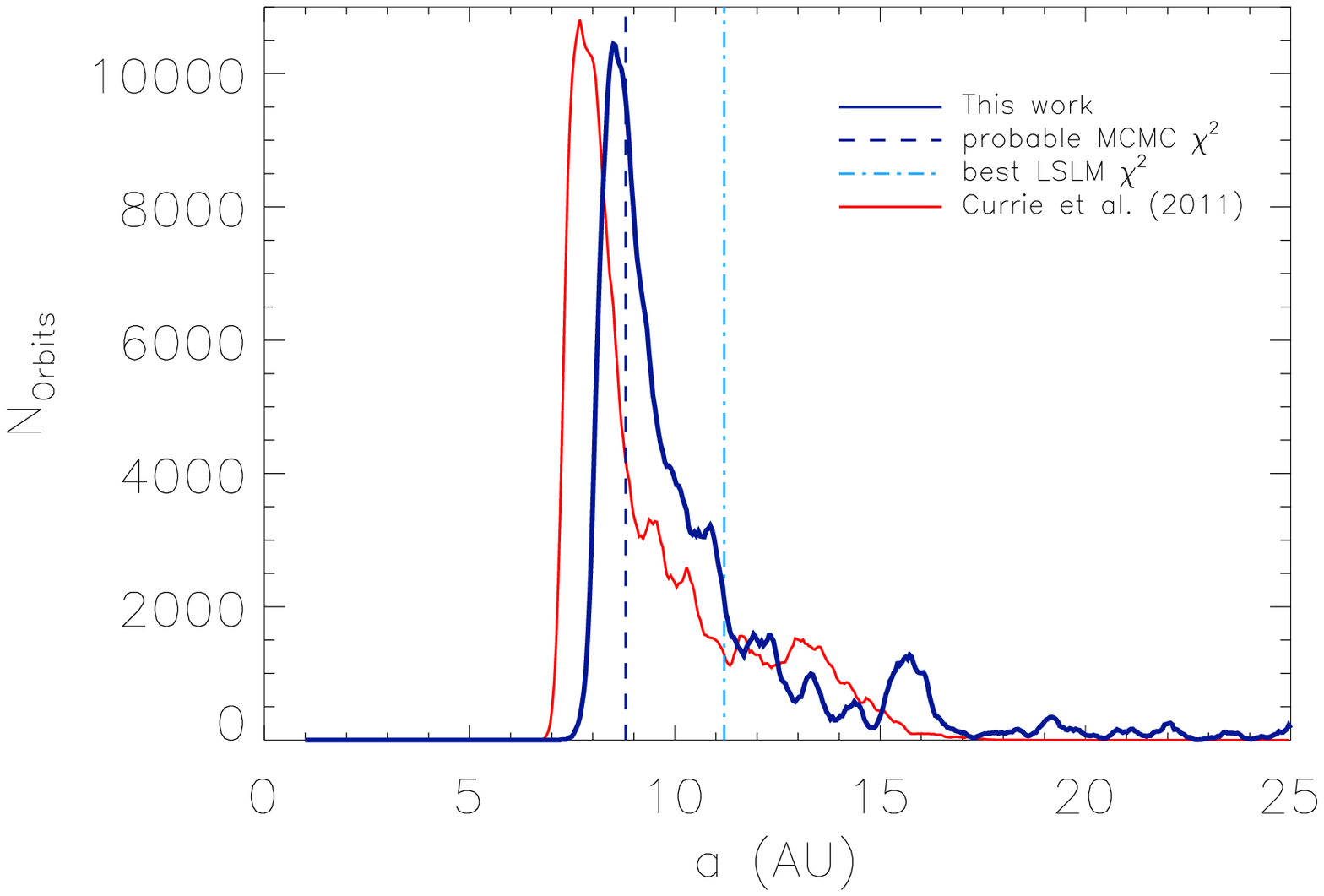}\hfil
\includegraphics[angle=0,width=0.33\textwidth]{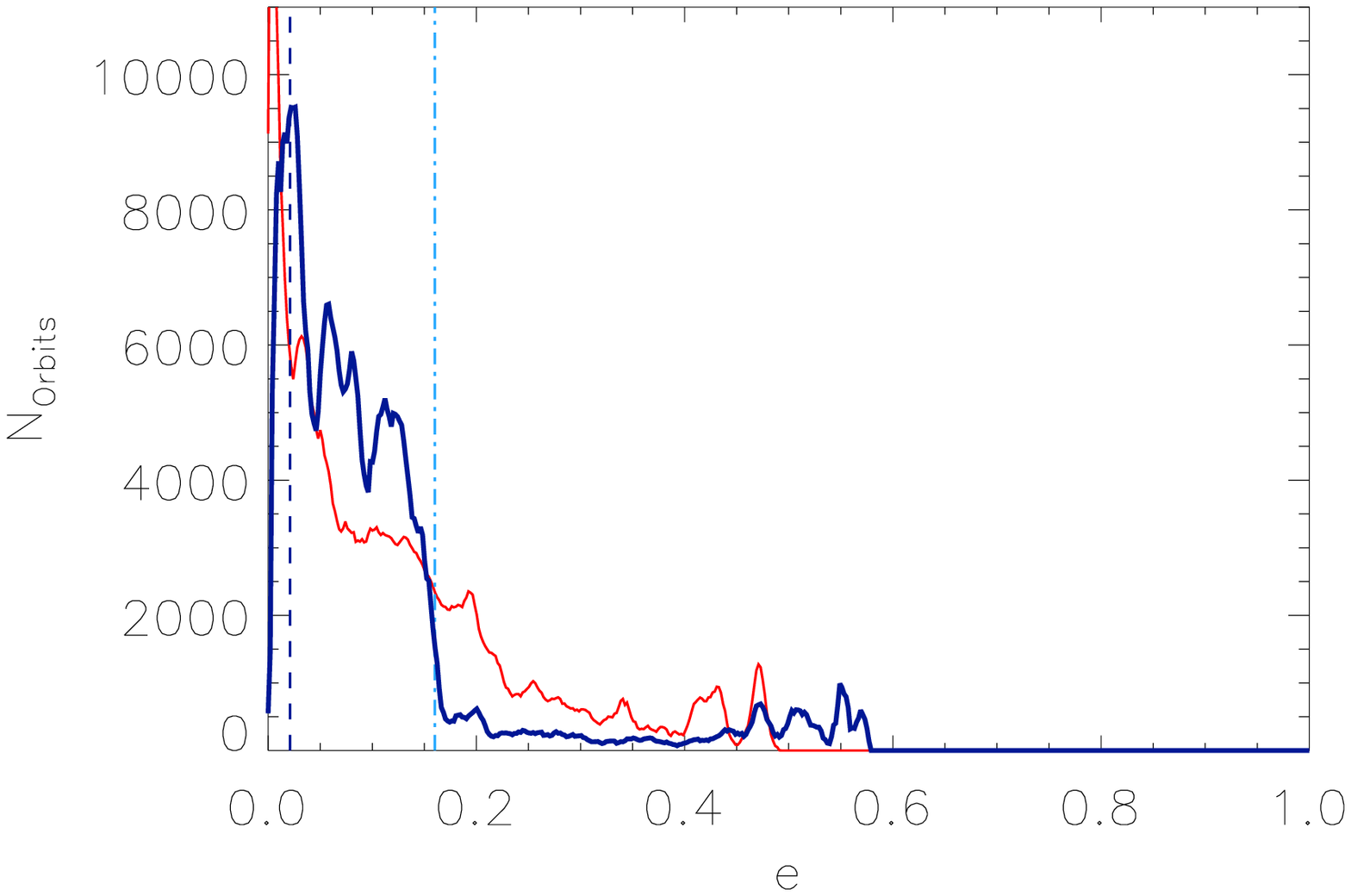}\hfil
\includegraphics[angle=0,width=0.33\textwidth]{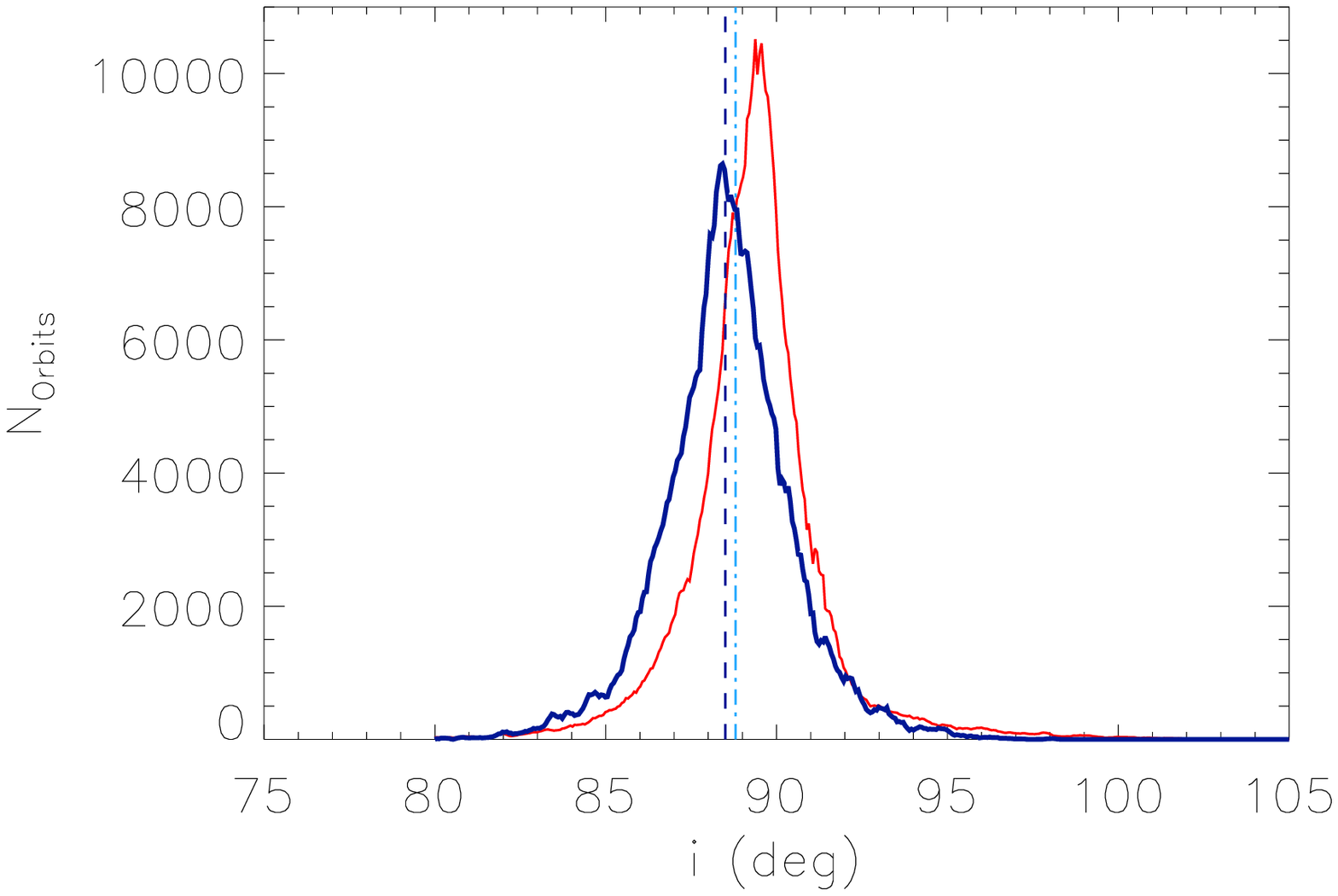}}
\makebox[\textwidth]{
\includegraphics[angle=0,width=0.33\textwidth]{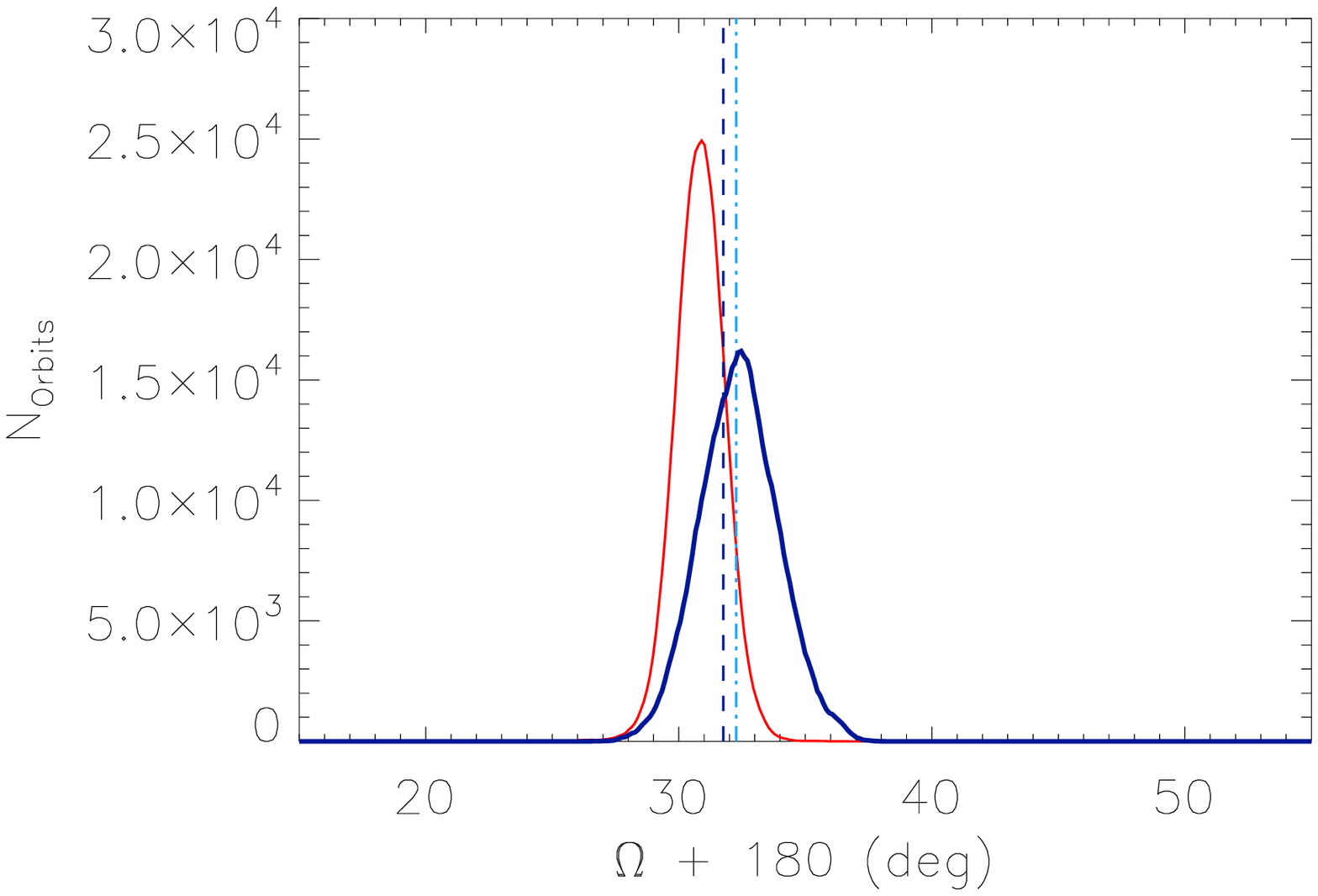}\hfil
\includegraphics[angle=0,width=0.33\textwidth]{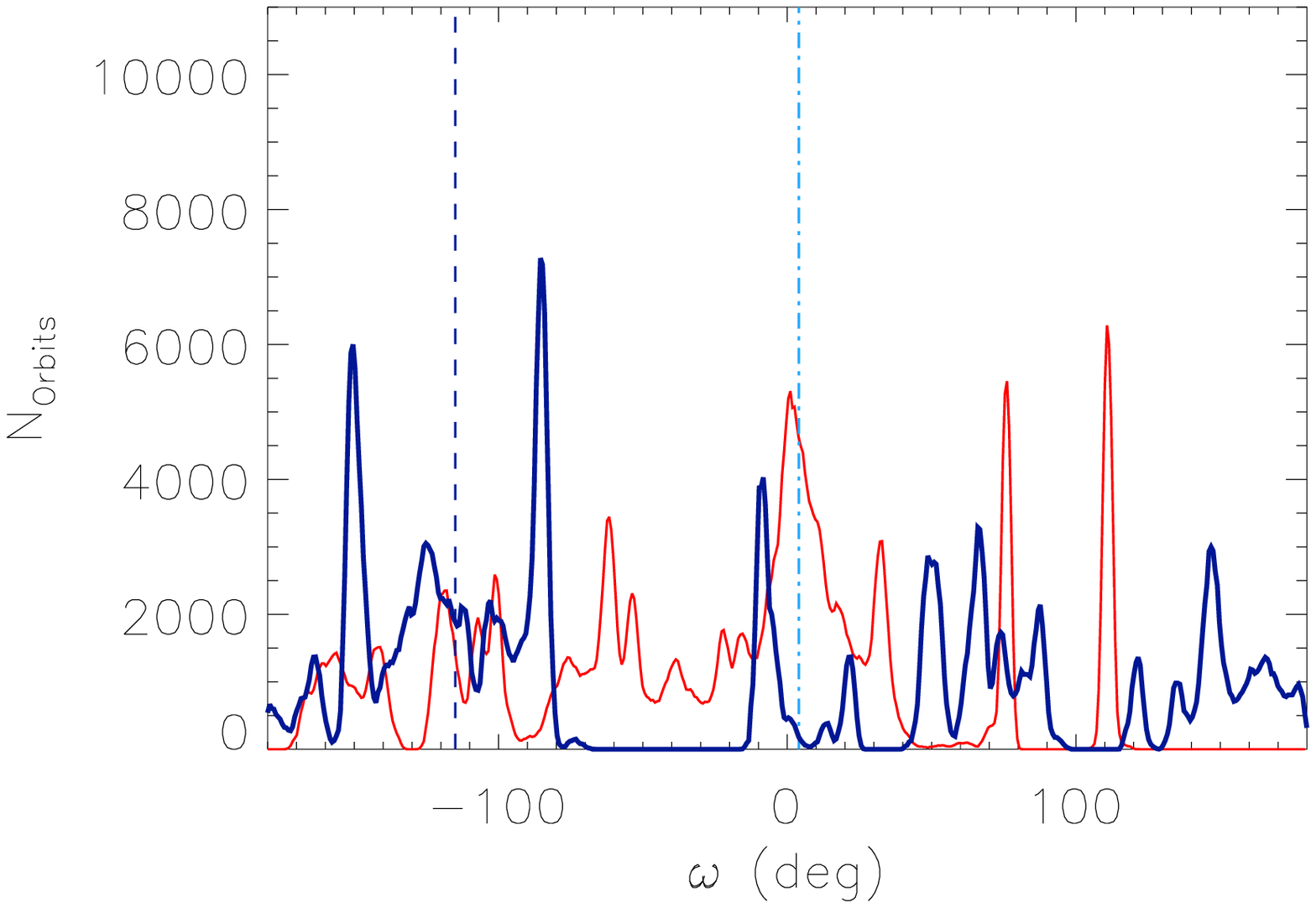}\hfil
\includegraphics[angle=0,width=0.33\textwidth]{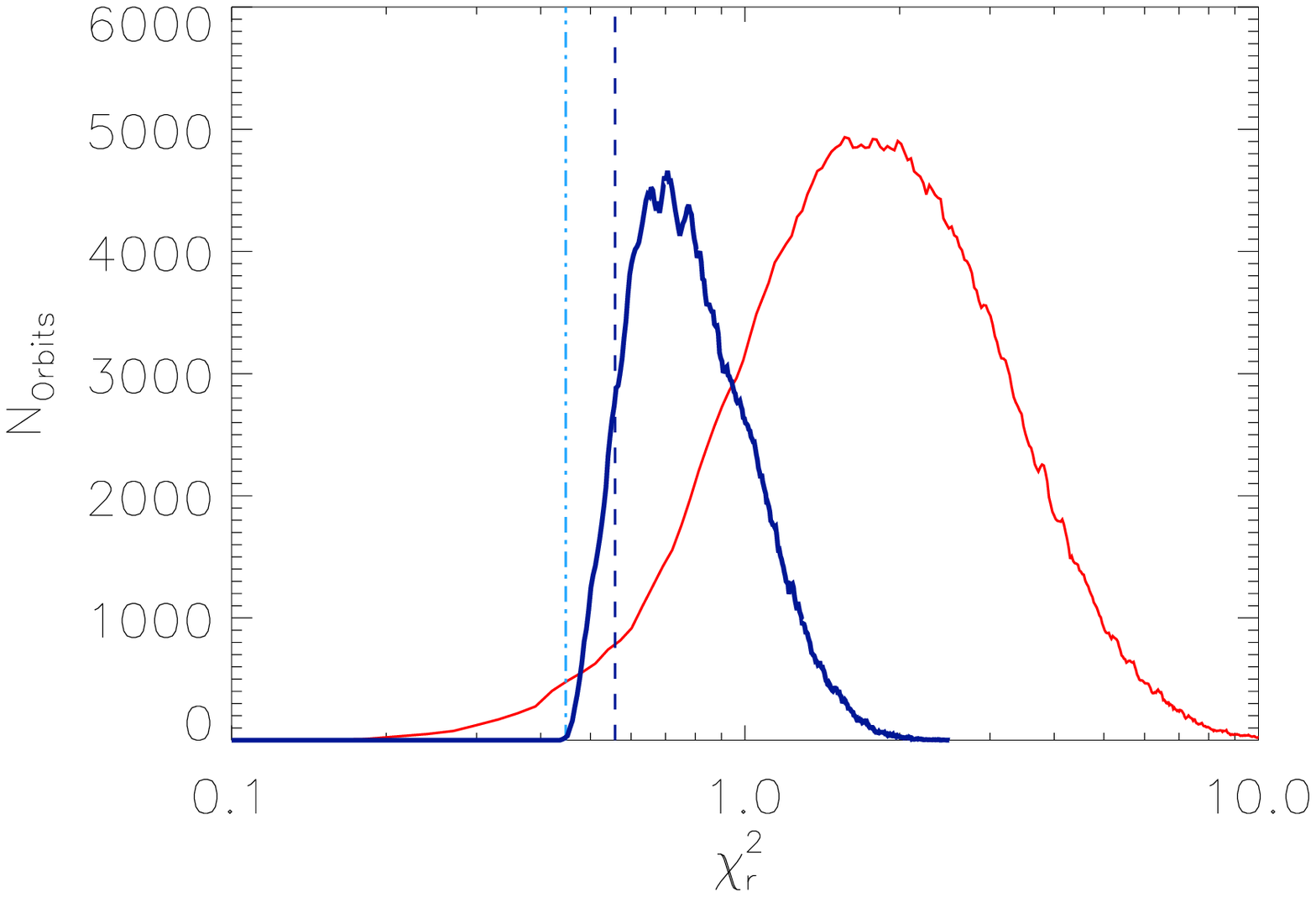}}

\caption{Results of the MCMC fit of the astrometric data of \bpb:
  statistical distribution of the orbital elements $a$ (top left), $e$
  (top middle), $i$ (top right), $\Omega$ (bottom left) and $\omega$
  (bottom middle). We also show the distribution of $\chi_r^2$ for the
  solutions obtained (bottom right).  In each plot, the \textit{blue}
  line show the results of our MCMC study. The \textit{dot-dashed}
  line indicates the position of the best-fit LSLM $\chi_r^2$ model
  obtained, and the \textit{--dashed} line shows the position of an
  example of highly probable orbits according to our MCMC study see
  Table~\ref{mle}). The results of our MCMC study using the
  dataset of \citet{cur11} are indicated by a \textit{red} line.}

\label{fig:mcmc}
\end{figure*}

A Keplerian model was then fitted to our $(\Delta\delta,\Delta\alpha)$
results of Table~\ref{tab:astro}, to derive the orbital period $P$ (or
equivalently the semi-major axis $a$, using the stellar mass
$M_*=1.75\,\msun$; Crifo et al. 1997), the $e$, $i$,
$\Omega$, $\omega$, and the time for periastron passage $t_p$.  To
constrain the \bpb's orbit, we used two complementary fitting methods:
a least squares Levenberg-Marquardt (LSLM) algorithm \citep{numrec} to
search for the model with the minimal reduced $\chi^2$, and a more
robust statistical approach using the Markov-chain Monte Carlo (MCMC)
Bayesian analysis technique \citep{ford05,ford06}. The details of the
MCMC analysis, which was adapted to this astrometric
characterization, are reported in Appendix A. The best-fit LSLM solution
found, and a typical example of a ``highly probable orbit'' according to
the result of the MCMC study, are given in Table~\ref{mle}. These
orbits are plotted in Fig.~\ref{orbplots}, in a plane containing the
line of sight, as well as the positions of the planet at various
observing dates. The results of both orbital fitting methods are also
shown for the five orbital elements $(a,e,i,\Omega,\omega)$ in
Fig.~\ref{fig:mcmc}. In that figure, we also display the $\chi^2_r$
distribution obtained for both analyses (considering a degree of
freedom $N-P=12$, where N is our number of measurements, here 18, and
P the number of parameters in our orbital model, here
6). Additional figures illustrating the goodness of both
  orbital fits are given in Appendix B.

When comparing the results of the best-fit LSLM solution with our MCMC
distributions, our first striking result is that they do not
match. The best-fit LSLM solution has a relatively good $\chi^2_r$
compared to the $\chi^2_r$ distribution of the MCMC. However, this
discrepancy seems to illustrate that our astrometric measurements are
still too sparse to derive a deep LSLM $\chi^2_r$ mininum, as the
region of probable orbits is probably fairly flat. Consequently, our
confidence in the LSLM approach is low, and a robust determination of
the orbital parameters of \bpb\ is currently impossible with this
fitting method. In contrast, we appear to have more success with the
statistical MCMC approach, which produces the probable ranges of
orbital elements of \bpb, enabling us to explore the significance of
each orbital solution.

The results of our MCMC analysis, reported in
  Fig.~\ref{fig:mcmc}, are distributions of the orbital parameters
  that are far from gaussian, except for the inclination, and the longitude of
  ascending node. The semi-major axis distribution peaks at
8.0--9.0~AU, and most eccentricities are given by $e \le 0.17$. The
inclination appears to be extremely concentrated close to $90\degr$, but
with a peak at $i = 88.5\pm1.7\degr$ (considering a
  confidence level at 1~$\sigma$). This is indicative of a $\sim
1.5\degr$ tilt with respect to a strict edge-on configuration. The
longitude of ascending node is also fairly well-constrained at $\Omega
= -147.5\pm1.5\degr$ ($\Omega + 180 = 32.6\pm1.5\degr$,
with a 1~$\sigma$ error). Finally, the
statistical distribution of the argument of periastron $\omega$ is
more erratic, and does not appear to have clear solutions, which is not
unsurprising owing to the low values of the eccentricity
distribution. \citet{cur11} performed a similar analysis, but on a
smaller dataset that had not been homogeneously processed, thus with presumably
larger systematics (different data processing, and possible lack of
NACO rotator offset calibration). With the same dataset, we
  found similar results to \citet{cur11}, using our own MCMC
  code. The results of both studies, with four epochs (Table~1 from
  Currie et al. 2011) and eight epochs (Table~2, this work), are compared
  in Fig.~\ref{fig:mcmc}.  To a first order, they give similar
results, although we note some differences. The peak of our semi-major
axis distribution is shifted by about 1~AU to larger values. Our
eccentricity distribution has a sharper cutoff that excludes a larger
fraction of high eccentricity solutions, thanks to our most recent
observations in 2010 and 2011.  Our distributions of inclination,
and longitude of ascending node are also shifted by respectively -1.0
and $1.5\degr$. Finally, the distribution of $\chi^2_r$ is better
constrained in our case ($\chi^2_r=0.75\pm0.30$, respectively to
$\chi^2_r=1.75\pm1.25$), corroborating a more concentrated and robust
set of orbital solutions. A peak value lower than unity however
indicates that our uncertainty estimates are probably
conservative. Further astrometric monitoring of \bpb, during the next
quadrature, will be mandatory to improve the planetary orbital
characterization.

\section{Discussion}

From our orbital fit analysis, four important outcomes arise: the
semi-major axis of \bpb\ falls in the probable range of 8--9.0~AU, the
eccentricity distribution is concentrated at $e\la 0.17$,
and the longitude of ascending node is fairly well-constrained at
$-147.5\pm1.5\degr$, as is the inclination distribution which peaks
at $88.5\pm1.7\degr$. The existence of a giant planet orbiting
\bp\ has been proposed by various studies during the past few decades
(e.g. Freistetter et al. 2007). The main indirect indicators are ({\it i}) the inner warped component of the \bp\ circumstellar disk,
together with additional asymmetries observed in the outer part
\citep{mou97,kal95}; ({\it ii}) the photometric transit-like event observed
in 1981 \citep{lec97}; and ({\it iii}) the cometary activity observed in the
absorption spectrum of \bp\ \citep{fer87,xxi,vid98,pet99}. We discuss below
how each of these observational findings may be related to both the existence and
the orbital and physical properties of \bpb.

\subsection{Disk - Planet configuration}

Dedicated scattered-light studies have accurately mapped the
morphology of the \bp\ disk \citep{kal95,hea00,gol06}. They revealed mainly
a nearly edge-on disk composed of a main disk observed beyond
80~AU, and an inner warped component (at less than 80~AU), inclined by
$2-5^o$ with respect to the main disk position angle. The simulations of
\citet{mou97}, \citet{aug01}, and more recently \citet{daw11}
demonstrated that the presence of a planet orbiting the star at $10\,$AU,
misaligned with the main disk, could actually form and sustain the
\bp\ inner warped disk. The observing challenge is then to test wether
\bpb\ might be this perturbating planet, that hence has with a
significantly inclined orbit with respect to the main disk
midplane. \citet{cur11} presented evidence of a misalignment
between the planet and the inner warped disk of \bp\, concluding that
the planet was orbiting inside the main disk's orbital plane. We,
however, do not confirm these results based on simultaneous
measurements of the planet, and the main disk positions. This work,
detailed in Lagrange et al. (2012), and using our $K_s$/S27
measurements of November 16, 2010 with a dedicated analysis for the
disk orientation, shows that the current location of \bpb\ is above
the midplane of the main disk (which has a position angle of
$\rm{PA}_{\rm{main-disk}}=209.0\pm0.3$~deg in the south west
direction). This position is more compatible with the warp component
orientation tilted by 3.5--4.0~deg (i.e with
$\rm{PA}_{\rm{warped-disk}}=212.5-213.0$~deg). Moreover, the current
distribution of longitude of ascending node $\Omega$, which peaks at
$-147.5\pm1.5\degr$ (and would correspond to a position angle of 32.5
or $212.5\degr$), can also be directly compared with the main disk and
the warp orientations. This distribution currently supports an orbital
plane for \bpb\ that does not coincide with the main disk midplane, but
more probably with the warp component. The hypothesis that the
\bpb\ planet is the one responsible for the inner warped
morphology of the \bp\ disk, without evoking any planetary
inclination damping (an alternative scenario proposed by
\citet{daw11}), remains valid. We may therefore still conclude
  that the \bpb\ planet continues to excite the disk of planetisimals, forcing
  them to precess about its misaligned orbit.

\subsection{1981 Transiting event}

\begin{figure}
\centerline{\includegraphics[angle=0,width=0.33\textwidth]{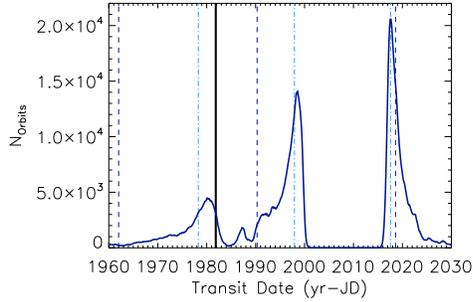}}
\caption{MCMC fit to the transit dates of \bpb\ in front of the line
  of sight. The plotting conventions are the same as in
  Fig.~\ref{fig:mcmc}. In addition, the date of the transiting
    event observed by \citet{lec97} is indicated by a thick vertical
  line.}
\label{fig:transits}
\end{figure}

\citet{lec95} reported significant photometric variations in November 1981 with a
peculiar transit-like event on November 10, 1981. \citet{lec97} showed
that a planet with 2--4 times the radius of Jupiter, orbiting at $\sim
9\,$AU at most could well be responsible for the photometric
variations they reported. \citet{lec09} investigated this issue on
the sole basis of the 2003 detection \citep{lag09} of \bpb. They found
that a transit of \bpb\ in November 1981 could be compatible with a
quadrature position in November 2003, assuming a semi-major axis in
the range [7.6--8.7]\,AU. They also predicted that if the planet
detected in 2003 on the NE side of the disk matched the one they
predicted, it should reappear on the SW side in 2009 roughly where it
was reobserved.  We reinvestigate this prediction based on
our MCMC orbital fit, considering both that we find a peak
for the inclination distribution at $88.5\pm1.7\degr$, and the transit
dates predicted in our set of orbital solutions. A tilt larger than
0.1~deg with respect to strictly edge-on configuration would simply
exclude the possibility that \bpb\ is transiting along the line of
sight. However, the sphere of influence of \bpb\ with a Hill radius of
about 1~AU (angular extension of about 7~deg at 8--9~AU) would still
cross the line of sight (even considering all MCMC
  solutions), and therefore influence the \bp\ photometry (if filled
with absorbing or scattering material).  Assuming this hypothesis,
we can still compare the date of the transit-like event of November
1981 with the predicted transit dates of the sphere of influence of
\bpb\ between 1960 and 2030 obtained by our MCMC analysis (see
Figure~\ref{fig:transits}). The MCMC distribution of transit dates
show that the parameters of the most recent ($\sim1999$) and next ($\sim2018$) transits
are somewhat well-constrained. The peak is broader in $\sim1980$. The
suggested transit date of November 1981 falls to the right edge of
that peak (although not at the center) and remains compatible with
the current set of orbital properties obtained from our MCMC analysis.

\subsection{The \bp\ cometary activity}
Transient redshifted spectral events have been regularly monitored in
the absorption spectrum of \bp\ \citep{fer87,xxi,vid98,pet99}, and
attributed to the sublimation of numerous star-grazing planetesimals
crossing the line of sight, which are referred to the falling evaporating
bodies (FEBs) phenomenon. Their origin has been tentatively related to
mean-motion resonances with a Jovian planet orbiting the star
\citep{bm96,bm00,pth01,bp3d}. Several constraints have been
deduced from dynamical studies of the FEBs scenario, suggesting that:
\begin{enumerate}
\item The planet responsible for the phenomenon is massive enough
  ($\sim$ jovian) to allow numerous bodies to be trapped in
  mean-motion resonances.
\item Its orbit is slightly eccentric ($e\ga 0.05$--0.1) to allow
  bodies trapped in the resonances to see their eccentricity pumped up
  \citep{bm96,bm00}.
\item The longitude of periastron $\varpi$ of the planet with respect to the
  line of sight is $\sim -70\degr\pm20\degr$ \citep{pth01}, to
  enable the statistics of the Doppler velocities of the FEB spectral
  signatures to match the observed ones which are largely strongly biased towards
  redshifts.
\item Finally, the planet location is no further away than $\sim 20\,$AU,
  otherwise the FEBs would hardly be able to get into the dust sublimation zone.
\end{enumerate}

The \bpb\ planet obviously has orbital and physical properties 
compatible with constraints 1/ and 4/. The situation is less
straighforward for the constraints 2/ and 3/. Eccentricities higher
than $\sim 0.05$--0.1 are indeed fully compatible with our fit, but
circular orbits cannot not excluded. Finally, the longitude of perisatron
$\varpi$ measured from the line of sight is related to the argument of
periastron $\omega$ in our fit. The parameter $\omega$ is measured from the $XOY$
plane of our reference frame, i.e., the plane of the sky. Assuming
an edge-on orientation of the disk, we then have
$\omega=\varpi+\pi/2$.  Thus, $\varpi\simeq-70\degr\pm20\degr$ means that
$\omega\simeq 20\degr\pm20\degr$. Unfortunately, our constraint on
$\omega$ remains too low to state whether condition 3/ is fulfilled
or not, partly due to our upper limit on the eccentricity distribution at
$e\la 0.17$). Further measurements are needed to yield conclusive results.

%

\section{Conclusion}

We have reported the results of our analysis of follow-up observations of the astrometric
positions of \bpb\ relative to $\beta$ Pic. Together with previously
archived data including available astrometric calibrations, we
have homogeneously reprocessed all relevant astrometric measurements of
\bpb\, at nine different observing epochs. We have taken into account the
various contributors to the uncertainty in our astrometric analysis,
including the determination of the star position in the low-flux part
of the saturated PSF, the influence of the PSF residuals on the
companion position, and the errors related to the detector
calibration, and the NACO Nasmyth-rotator position. We then used
orbital-fitting techniques to derive the most probable orbital
solutions for the \bpb\ planet: a least squares Levenberg-Marquardt
algorithm and a Markov-chain Monte Carlo Bayesian analysis. As our
measurements do not cover the complete planetary orbit, and are biased
towards the most recent epochs since the planet recovery, the
Markov-Chain Monte Carlo approach provides more robust and reliable
orbital solutions for $\beta$ Pic\,b.  The most probable ones favor a
low-eccentricity orbit $e\la0.17$, with a semi-major axis of
8--9\,AU corresponding to orbital periods of 17--21\,yrs, an
inclination that peaks at $88.5\pm1.7\degr$, and a longitude of
ascending node fairly well-constrained at $-147.5\pm1.5\degr$. Our
results support the previous astrometric studies of \citet{lag10} and
\citet{cur11}, although our present study have indeed provided a more robust
set of orbital solutions. Our conclusions support the idea that the
planet is not in the midplane of the main disk. The planet's position
is more compatible with the warped component orientation, thus
corroborating the idea that \bpb\ is responsible for the inner warped disk
morphology. The current orbital solution of \bpb\ is still consistent
with the sphere of influence of the planet being responsible for the
1981 transiting event, although a deviation of more than 0.1~deg from
an edge-on configuration would exclude a planetary transit. Finally,
the planet's predicted mass, eccentricity, semi-major axis, and
longitude of periastron also imply that \bpb\ is likely to be the origin of the
cometary activity observed in the \bp\ system.  Further deep imaging
characterization should help us to more tightly constrain the orbital parameter space,
once the planet has passed the quadrature (most probably in
2013). Further spectroscopic or multi-photometric observations should
also help us to determine the underlying physics of this giant planet in the
framework of current planetary atmosphere studies (Bonnefoy et
al. 2010; Janson et al. 2010; Skemer et al. 2011, Madhusudhan et
al. 2011; Barman et al. 2011a, 2011b). Therefore, much is to be
expected from future extreme-AO instruments SCExAO/Subaru (Guyon
2010), GPI/Gemini (Macintosh et al. 2008), SPHERE/VLT (Beuzit et
al. 2008) in the coming years.


\bibliographystyle{aa}

\begin{acknowledgements}

We would like to thank the staff of ESO-VLT for their support at the
telescope. This publication has made use of the SIMBAD and VizieR
database operated at CDS, Strasbourg, France. Finally, we
acknowledge support from the French National Research Agency (ANR)
through project grant ANR10-BLANC0504-01 and the {\sl Programmes
  Nationaux de Plan\'etologie et de Physique Stellaire} (PNP
\&\ PNPS), in France.

\end{acknowledgements}

\Online

\begin{appendix} 

\section{Convention and Markov Chain Monte Carlo adapted for Astrometry}

The astrometric position of the planet relative to the star is
described by Eqs.~(\ref{xmodel}) and (\ref{ymodel}). Fo any orbital
solution, $\Omega$ and $\omega$ changed to $\Omega+\pi$, $\omega+\pi$,
$v+\pi$ respectively yields the same astrometric data. In the context
of $\Omega$, the difference between $\Omega$ and $\Omega+\pi$ actually
resides in the $z$-motion (along the line of sight). If we consider a
nearly edge-on orbit, we expect the longitude of the ascending node
$\Omega$ to match the PA of the astrometric data (here, $\Omega \simeq
34\degr$ or $\Omega \simeq 211 = -149\degr$ if we consider an edge-on
orbit for $\beta$ Pic~b). By convention, $\Omega$ must thus
corresponds to the PA when the $z$-coordinate grows, i.e., when the
planet is moving towards the observer. In the context of $\beta$ Pic,
the rotation sense of the gaseous disk was determined by
\citet{olo01}: the NE branch is receding from us while the SW branch
is approaching. If we assume that the planet is moving in the same
sense as the disk, then \bpb\ was receding in the 2003 observations
and is approching now, and it passed behind the star in between. This
means that the ascending node is located towards the SW branch of the
orbit, or that $\Omega\simeq-149\degr$. We will use this property to
distinguish between solutions that yield the same astrometric
data. Note that this determination occurs in any case after the
fitting procedure. In both approaches, we fit $\omega+\Omega$ and
$\omega-\Omega$. Eqs.~(\ref{xmodel}) and (\ref{ymodel}) can be
rewritten unambiguously as a function of $\omega+\Omega$ and
$\omega-\Omega$).

In the context of the Markov Chain Monte Carlo, let us briefly recall
the detail of that technique that we here adapted for $\beta$
Pic~b. Let us call $\vec{x}$ the model parameters vector we want to
constrain and $\vec{d}$ the data vector.  We want to determine the
posterior distribution $p(\vec{x}|\vec{d})$, i.e., the probability
function of the parameter vector $\vec{x}$ given the data vector
$\vec{d}$ and its error vector. This requires the knowledge of a prior
probability function $p_0(\vec{x})$ for $\vec{x}$.  A Markov chain is
a sequence of successive set of trial values $\vec{x}_i$ ($i\geq 1$)
for $\vec{x}$.  The Metropolis-Hastings (M-H) algorithm \citep{ford05}
is used to derive $\vec{x}_{i+1}$ from $\vec{x}_i$ via a transition
probability.  After convergence, the equilibrium distribution of the
chain equals the posterior distribution
$p(\vec{x}|\vec{d})$. \citet{ford06} suggests several assumptions for
MCMC adapted to the search of exoplanets by radial velocity, depending
on the kind of orbit we are looking for. We adapt here his
recommendations to our astrometric fit.  Following \citet{ford06}, we
assume the prior distribution $p_0(\vec{x})$ to be uniform in
$\vec{x}=(\log P, e, \cos i,\Omega+\omega,\omega-\Omega, t_p)$.
However, the work is done on the parameter vector
$\vec{u(x)}$. 

 Our main motivation in using $\vec{u}$ as a
  variable instead of $\vec{x}$ was to improve the convergence of the
  Markov chains, as suggested by \citet{ford06}. A good way to achieve
  this is to avoid singularities. For instance, using
  $e.cos(\omega+\Omega)$ and $e.sin(\omega+\Omega)$ instead of
  $\omega$ and $\Omega$ removes the non-regularity at $e=0$ ($\Omega$
  is not defined); using $\omega+\Omega$ causes these variable to be
  still well defined even if $i=0$, which is not the case for $\omega$
  and $\Omega$ individually. Finally, dividing by $\sqrt(1-e^2)$
  avoids to test non-physical values at large eccentricities. 

\vspace{0.1cm}
The following equation is then used:

\begin{eqnarray}
\vec{u}(x) & = & \left(\frac{\cos(\omega+\Omega+v_0)}{P},
\frac{\sin(\omega+\Omega+v_0)}{P},\frac{e\cos(\omega+\Omega)}{\sqrt{1-e^2}},
\right.\nonumber\\
&&\frac{e\sin(\omega+\Omega)}{\sqrt{1-e^2}},
(1-e^2)^{1/4}\sin\frac{i}{2}\cos(\omega-\Omega),\nonumber\\
&&\left.(1-e^2)^{1/4}\sin\frac{i}{2}\sin(\omega-\Omega)\right)
\end{eqnarray}
where $v_0$ is the true anomaly of the planet at a specific better
constraining date $t_0$. Here we fix $t_0$ to be the date of the 2003
observation (Nov. 13, 2003). We run 10 chains in parallel and use the
Gelman-Rubin statistics as convergence criterion \citep[see details
  in][]{ford06}.  The results of the MCMC runs are reported in the
context of \bpb\ in Fig.~3 and 4.

\section{Orbital fit material}

\begin{figure}[h]
\includegraphics[angle=0,width=\columnwidth]{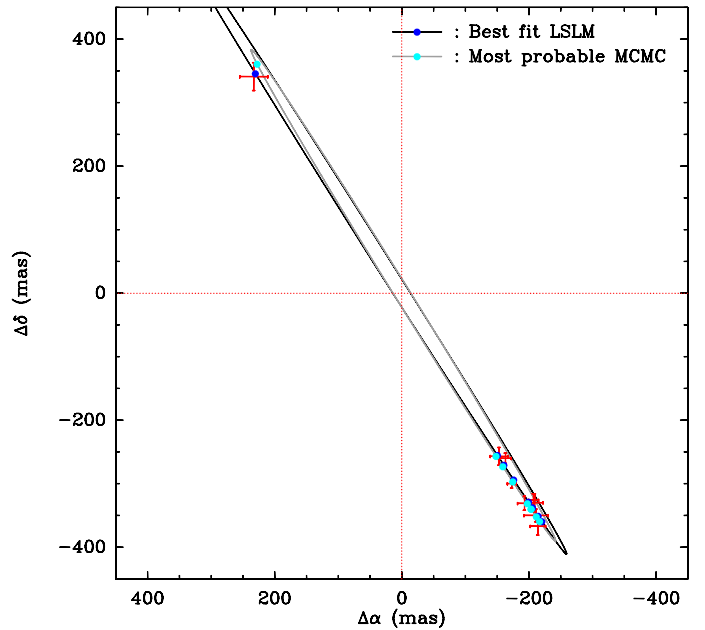}
\caption{Measured astrometric positions of \bpb\ relative from A on the plane of sky used in the
  present work to characterize the orbital parameters of the
  \bpb\ planet. Together with the observed measurements are
  overplotted the orbital solution the best LSLM $\chi^2$ model and
  the highly probable orbit obtained with the MCMC approach.}
\label{fig:added1}
\end{figure}

\begin{figure}[h]
\includegraphics[angle=-90,width=\columnwidth]{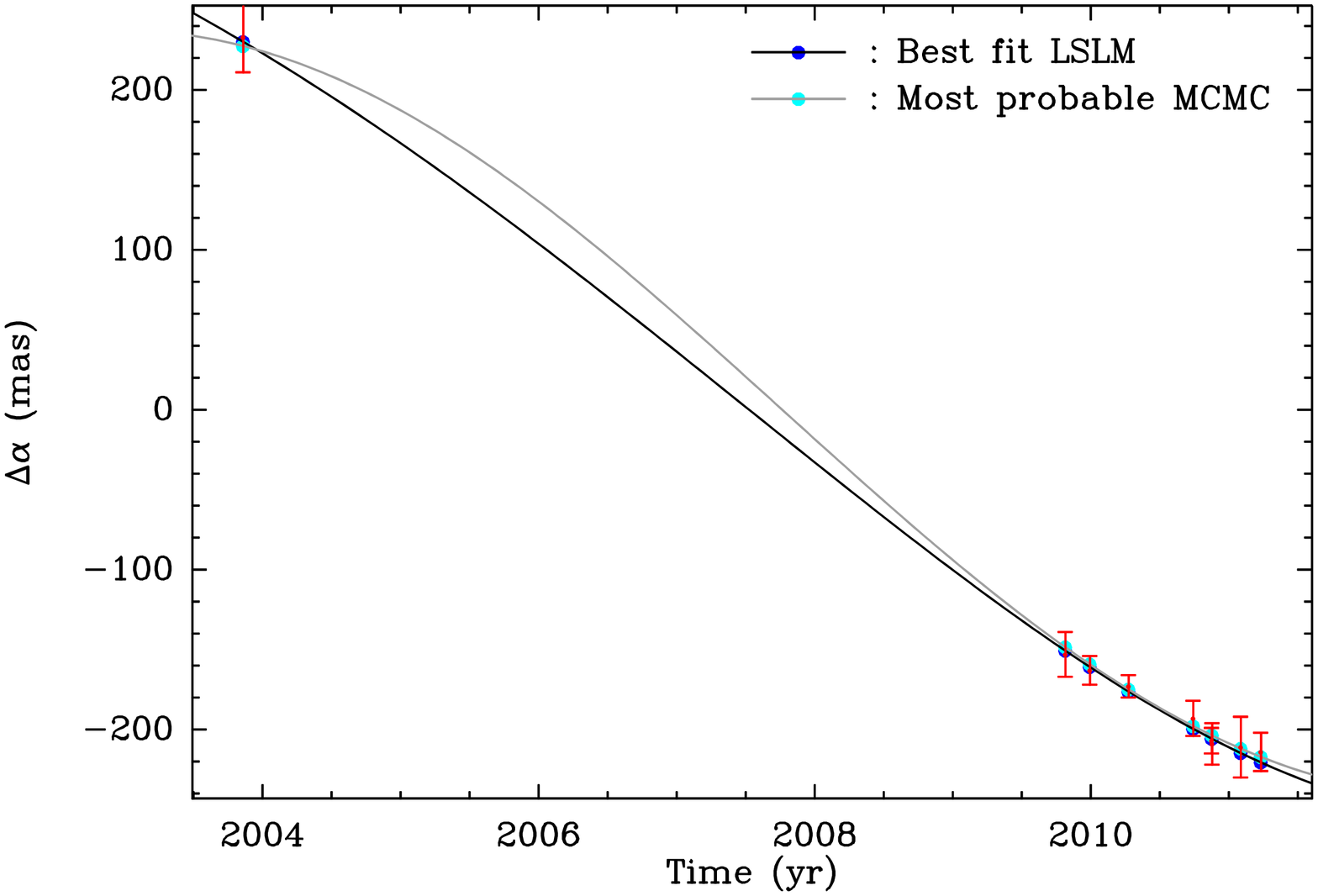}
\caption{Measured astrometric positions of \bpb\ relative from A in
  $\Delta\alpha$ as a function of time. Both orbital solutions of  the best LSLM $\chi^2$ model and
  the highly probable orbit obtained with the MCMC approach are overplotted.}
\label{fig:added2}
\end{figure}

\begin{figure}[h]
\includegraphics[angle=-90,width=\columnwidth]{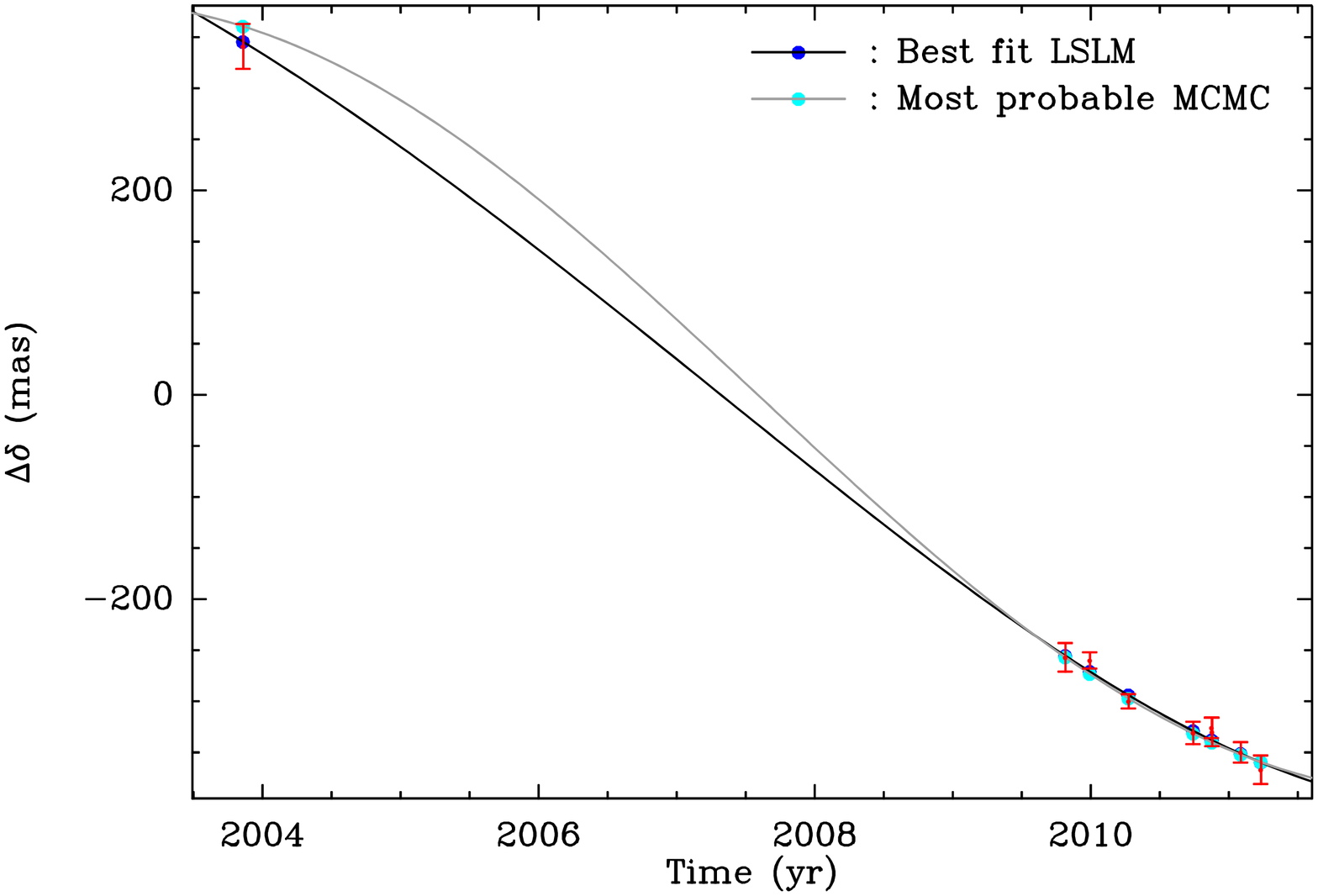}
\caption{Measured astrometric positions of \bpb\ relative from A in
  $\Delta\delta$ as a function of time. Both orbital solutions of  the best LSLM $\chi^2$ model and
  the highly probable orbit obtained with the MCMC approach are overplotted.}
\label{fig:added3}
\end{figure}

You will find below the astrometric data of \bpb (with their error
bars from Table~\ref{tab:astro}), together with the results of the
orbital fit for: 1/ the best LSLM solution, and 2/ the "highly
probable orbit'' according to the MCMC approach. The details of the
orbital parameters of both solutions are given in
Table~\ref{mle}). Fig.~\ref{fig:added1} gives the results of the
orbital fit on the projected sky plane.  Fig.~\ref{fig:added2}
and~\ref{fig:added3} give the evolution of the relative astrometric
values of $\Delta\alpha$ and $\Delta\delta$ as a function of time.

\end{appendix}

%

\end{document}